\documentclass{ws-ijmpa}
\usepackage[super,compress]{cite}
\usepackage{graphicx}
\usepackage{amsmath}
\usepackage{hyperref}
\usepackage{xcolor}
\begin{document}
\markboth{D. Martínez C, M. de la Cruz, S. Patiño, L. Z\'uñiga}{Implications of Seiberg-Witten map on type-I superconductors}

\catchline{}{}{}{}{}
%
%%%%%%%%%%%%%%%%%%%%%%%%%%%%%%%%%%%%%%%%%%%%%%%%%%%%%%%%%%%%%%%%%%%%

\title{\bf{Implications of Seiberg-Witten map on type-I superconductors}}

\author{Daniel Mart\'inez-Carbajal$^\dag$}
\address{Tecnológico de Estudios Superiores del Oriente del Estado de México, Paraje San Isidro s/n, col. Barrio de Tecamachalco,
	A.P. 41, Estado de México C.P. 56400, México\\ $^\dag$daniel.mc@otemexico.tecnm.mx}
 \address{Departamento de Física de Altas Energías, Instituto de Ciencias Nucleares, Universidad Nacional Autónoma de México, Apartado Postal 70-543, CDMX 04510, México\\
$^\dag$daniel.martinez.carbajal@correo.nucleares.unam.mx}

\author{Manuel de la Cruz$^\S$}
\address{Instituto de Física, Benemérita Universidad Autónoma de Puebla, Apartado Postal J-48, CP 72570, Puebla, Puebla, México\\
$^\S$mdelacruz@ifuap.buap.mx}

\author{Sergio Patiño-López$^\ddagger$}
\address{Departamento de Física de Altas Energías, Instituto de Ciencias Nucleares, Universidad Nacional Autónoma de México, Apartado Postal 70-543, CDMX 04510, México\\
$^\ddagger$sergio.patino@correo.nucleares.unam.mx}

\author{Leonardo D. Herrera-Zúñiga$^{*}$}
\address{Tecnológico de Estudios Superiores de Huixquilucan, Paraje El Río s/n, Col. La Magdalena Chichicaspa. CP 52773, Estado de México, México\\
$^{*}$sub.estudiosprof@huixquilucan.edu.mx}

\maketitle

\begin{history}
\received{Day Month Year}
\revised{Day Month Year}
\end{history}

\begin{abstract}
Guided by Pippard superconductivity, we incorporate the Seiberg-Witten map in the classical London theory of type-I superconductors when an external magnetic field is applied. After defining the noncommutative Maxwell potentials, we derive the London equation for the supercurrent as a function of the noncommutative parameter, up to second order in gauge fields expansions. Keeping track of the effects of noncommutative gauge fields, we argue that noncommutative magnetic field effects can be cast in the London penetration length similarly to non-local Pippard superconductivity. Also, we show that the flux quantization remains consistent relative to the commutative case. Our effective London penetration length reduces to the standard one in the commutative limit. These results allow us to argue that the framewor of noncommutative electrodynamics can give some insights into the anisotropic and non-local structure of superconductivity and Condensed Matter Systems, perhaps out of the ultra-microscopic scale regime.

%\textbf{Keywords}: Type-I superconductors, Seiberg-Witten map, London equation.
\keywords{Type-I superconductors; Seiberg-Witten map; London equation.}
\end{abstract}

%\tableofcontents
\section{Introduction and summary}

In the early years since the discovery of superconducting materials by H. K. Onnes (1911) and the development of a theoretical framework to describe them\footnote{Particularly, the Meissner-Ochsenfeld effect.}, many experimental setups have shown that the supercurrent equation, proposed originally by London brothers \cite{London-1935}, needs to be generalized due to deviations from the theoretical relation between current and Maxwell potential. An important effort was realized in \cite{pippard}, where the author promotes the supercurrent local equation to long-range order interactions between electrons. Pippard obtained a non-local expression of London current that was in good agreement for some doped materials (dirty limit). Towards a modification of penetration length parameter $\lambda_{L}$, another important contribution was the introduction of the coherence length $\xi$, related to the mean free path of electrons and the Copper pair. Nowadays, the relation $\lambda_{L}/\xi$ allows us to classify materials in type-I and type-II superconductors, being $\xi$ estimated using uncertainty-principle arguments \cite{1950RSPSA.203..210P, book:17888}
\begin{equation}
    \Delta x\geq \dfrac{\hbar}{\Delta p}\approx \dfrac{\hbar v_{F}}{K_{B}T_{c}}\equiv a\xi_{0},\label{PipEst}
\end{equation}
where $v_{F}$ is the Fermi velocity. Pippard relates the above expression with the correlation length of electrons and, using experimental data, determined that $a=0.15$; close to the value obtained in BCS framework ($a=0.18$)~\cite{PhysRev.108.1175}. Therefore, the non-local effects of Pippard theory can be interpreted as a modification of the effective size of the Cooper pair. In other words, the current $\textbf{J}(\textbf{r})$ depends on $\textbf{B}(\textbf{r}')$ throughout a volume of radius $|\textbf{r}-\textbf{r}'|\sim \xi_{0}$\footnote{There are two Pippard correlation lengths, $\xi$ and $\xi_{0}$ related with the electron mean free path $l$ $$1/\xi=1/ l+1/\xi_{0}.$$ At zero temperature $\xi_{0}=\xi_{GL}$ being the latter, the Ginzburg-Landau correlation length.}.

In this line of non-local thinking, one main question in modern Condensed Matter Theory is to what extent entanglement, coherence, non-locality and strong coupling are related to each other towards the complete comprehension of superconducting state? The non-local phenomenon is originated by the coherent condensation of electron pairs into a macroscopic quantum state, hence the characteristic distance of condensation is coherence length $\xi$. Therefore, this quantity represents a measure of the \emph{intrinsic} non-local nature of the superconducting state. For Type-I superconductors, in which $\xi\gg\lambda_{L}$, the Electrodynamics is non-local causing, among other effects, that the decay function of the fields within the superconductor is not exactly exponential. Nevertheless, it has been demonstrated that London theory is a good approximation for Type-I superconductors when substituting $\lambda_{L}$ by an effective penetration depth. \cite{1950RSPSA.203..210P,alexandrov2003theory}.

As we review in the next paragraphs, noncommutative (NC) field theories can be viewed as a mechanism to spread out the sources to promote non-locality field interactions due to a new kind of \emph{uncertainty} $\Delta x\Delta y\sim \theta$, where $\theta$ is the noncommutative parameter that allows us to control the effects of noncommutative geometry using well-posed expansions by virtue of the \emph{Seiberg-Witten} (SW) map.

In NC theories, our main motivation in this work, the space-time is discretized at Planck scale via the canonical commutation relations of space-time operators $\left[\hat{x}^{\mu}, \hat{x}^{\nu}\right]=i\theta^{\mu\nu}$ inducing an effective non-local interaction between the fields living in this Lorentz-discrete invariant background\footnote{See \cite{Hinchliffe:2002km} and the references therein for a review of these topics.}. A further result in NC field theory was in the context of String Theory \cite{Seiberg:1999vs}. These authors prove the existence of noncommutative geometry of gauge fields on supersymmetric $D$-branes and develop perturbation theory with noncommutative parameter $\theta$ for the gauge fields to all orders in $\theta$-expansions, known as SW map. An important feature of this theoretical tool is that we can rewrite the usual electrodynamics in terms of noncommutative counterparts to get a better understanding of the singularities and non-symmetric sources that arise in classical electrodynamics. Another property of SW map is the compatibility with full local gauge symmetry. All the above properties of SW map are relevant in this work.

We restrict ourselves to the study of type-I materials at zero temperature; where quantum fluctuations are expected to be important relative to the thermal ones, and we analyze NC effects in this regime. Hence, inspired by Pippard superconductivity, this work considers a SW $\theta$-expansion of the Maxwell fields in order to explore how physical penetration length of the superconductor is being affected. If the correlation length in Eq.~(\ref{PipEst}) has the same value in any coordinate (anisotropic-like behavior), then, by dimensional analysis $\theta\sim a^{2}\xi_{0}^{2}$. Therefore, the introduction of non-local NC electrodynamics would be justified at least in this NC scale\footnote{NC could give real physical predictions at energies $\sim$TeV.}. Moreover, $\xi_{0}\sim 1/\delta(0)$ at zero temperature hence, roughly speaking, NC intrinsically potentially changes the energy gap $\delta(0)$\footnote{We do not explore this sentence further because a complete NC Ginzburg-Landau theory is needed.}. At this respect, one of the main features of Pippard Superconductivity is that the relation between current and Maxwell potential changes by an effective $\lambda_{P}$ relative to that of London equation $\lambda_{L}$ thus, in this work we use $\theta$-expansion at $T=0$ to obtain effective London penetration length. Finally, the SW formalism has well posed formalism in classical Electrodynamics and we argue that these NC effects can give new insights to understand some features of superconductors or in general, Condensed Matter Systems. 

Section (\ref{sec2}) introduces the basic notion of noncommutative electrodynamics and Seiberg-Witten map that we use in this work. Section (\ref{sec3}) shows the general non-commutative version of London theory. Section (\ref{secc:8}) is dedicated to show explicitly the NC effective penetration length and the magnetic field corrections in $\theta$-expansions. Section (\ref{secc:4}) is dedicated to discussing the flux quantization in London theory in the context of the noncommutative framework and we show the $\theta$-expansions to all orders for the noncommutative gauge parameter $\widehat{\varLambda}$. Conclusion, discussion and further developments about NC effects are presented in (\ref{conclusiones}). An appendix is included related to noncommutative electrodynamics. Throughout the paper, we work in 4-dimensional mostly plus Minkowski space-time.

%%%%%%%%%%%%%%%%%%%%%%%%%%%%%%%%%%%%%%%%%%%%%%%%%%%%%%%%%%%%%%%%%%%%%%%%%%%%%%%%%%%%%%%%%%
\section{Noncommutative Gauge Theories}\label{sec2}
The celebrated London equation of superconductivity is derived from classical Electrodynamics with a set of novel assumptions \cite{London-1935}. On the other hand, in High Energy Physics, the noncommutative structure of space-time can be incorporated in Electrodynamics as an effective field theory, particularly on the behaviour of the electric and magnetic potentials. This is the approach that we consider in this work in order to explore the consequences of noncommutative effects on London superconductivity. To achieve this aim, we need to review the necessary NC theory that will be helpful for the next sections (see \cite{Hinchliffe:2002km, Seiberg:1999vs,wess2001non,Maceda:2016ety} for details).

The introduction of the \emph{Moyal-Groenewold bracket} $[A,\,B]_* :=A*B-B*A$ in gauge theory leads to regard the NC version of the field strength
\begin{equation}
\widehat F_{\mu\nu} := \partial_{\mu}\hat A_{\nu}-\partial_{\nu}\hat A_{\mu}-i[\hat A_{\mu},\,\hat A_{\nu}]_{*},
\label{eq:the field strength}
\end{equation}
which is compatible with the infinitesimal transformation law
$\ensuremath{\delta_{\hat{\varLambda}}\widehat{F}_{\mu\nu}=i[\hat{\varLambda},\,\widehat{F}_{\mu\nu}]_{*}}$, where $\hat{\Lambda}$ stands for NC gauge parameter that generalize the standard $\lambda$ in commutative gauge theory\footnote{Recall that the gauge parameter for a group in gauge theory is defined as $\lambda=\lambda_{a}T^{a}$ where $T^{a}$ are the generators of the associated gauge symmetry group that satisfy $\left[T^{a},T^{b}\right]=if^{c}T_{ab}$ and $\lambda_{a}$ are functions of the space-time coordinates. In $U(1)$ gauge theory, $\lambda (x)$ is the phase function of some field $\Phi$ associated with the gauge invariant transformation $A'\mapsto A+\partial_{\mu}\lambda$. The infinitesimal transformations are $\delta_{\lambda}F_{\mu\nu}=i\left[\lambda,F_{\mu\nu}\right]$ and $\delta_{\lambda}\Phi=i\lambda\Phi$.}.
We use the circumflex symbol \emph{hat} to distinguish the NC field from the standard one. On the other hand, for a generic field $\hat{\Phi}$ coupled to gauge field, the connection for the covariant derivative is written as 
\begin{equation} D_{\mu}\hat\Phi=\partial_{\mu}\hat\Phi-i\hat A_{\mu}*\hat\Phi,\label{eq:covariant D nc} \end{equation} which is equal to the standard covariant derivative up to the Moyal product.  Eq.~(\ref{eq:covariant D nc}) is compatible with the gauge transformation
\begin{equation}
\hat A_{\mu}\rightarrow \hat A_{\mu}^{\prime}=U* \hat A_{\mu}* U^{-1}+iU*\partial_{\mu}U^{-1},
\label{eq:gauge transformation-1}
\end{equation}
where $U$ is an element of the noncommutative group $U_{*}(1)=e^{i\hat{\varLambda}(x)}$. A novel feature of this construction is that, for any symmetry gauge group, the noncommutative effects due to Moyal-Groenewold bracket leave the action invariant $\delta_{\lambda}S=\widehat{\delta}_{\widehat{\varLambda}}S=0$. As pointed out in the introduction section, this effective NC Electrodynamics also arises from String Theory \cite{Seiberg:1999vs}, where the NC effects can be controlled using perturbative expansions with a $\theta$-parameter, maintaining the gauge symmetry intact.

In the framework of \cite{Seiberg:1999vs}, the potential, field strength and gauge parameter are considered to have the functional dependency 
\begin{equation}
\widehat{A}_{\mu}=\widehat{A}_{\mu}\left(A;\,\theta\right),\quad\widehat{F}_{\mu\nu}=\widehat{F}_{\mu\nu}\left(A;\,\theta\right),\quad\widehat{\varLambda}=\widehat{\varLambda}_{\lambda}\left(\lambda,\,A;\,\theta\right).
\end{equation}
Note that the NC gauge parameter depends on gauge fields. This fact makes the SW mappings more difficult but otherwise tractable in terms of $\theta$ expansions.
Using an iteratively procedure, the NC fields can be written as a power series in the noncommutative $\theta$-parameter~\cite{Ulker:2007fm}. Guided by the previous work \cite{Maceda:2016ety}, the field and gauge parameter can be written as a power series
%In order to solve iteratively the Eq.~(\ref{eq: noncommutative gauge transformations A}) and Eq.~(\ref{eq: noncommutative consistency equation}), 
\begin{equation}
\widehat{A}_{\mu}=A_{\mu}^{0}+A_{\mu}^{1}+A_{\mu}^{2}+\mathcal{O}^{n}(\theta),
\label{eq:noncommutative A}
\end{equation}
\begin{equation}
\widehat{\varLambda}_{\lambda}=\lambda+\varLambda_{\lambda}^{1}+\varLambda_{\lambda}^{2}+\mathcal{O}^{n}(\theta).\label{eq:gaugepar}
\end{equation}
The general form of the first order expansion of gauge potential and the associated generators were obtained in~\cite{Seiberg:1999vs} and read
\begin{eqnarray}
\varLambda_{\lambda}^{1} & = & \frac{1}{4}\theta^{\mu\nu}\{\partial_{\mu}\lambda,\:A_{\nu}^0 \},\label{eq:parameter first order}
\\[4pt]
A_{\gamma}^{1} & = & -\frac{1}{4}\theta^{\mu\nu}\{A^0_{\mu},\:\partial_{\nu}A^0_{\gamma}+F^0_{\nu\gamma}\},
\label{eq:SW Potential A first order}
\end{eqnarray}
where $\{A,B\}_* :=A*B+B*A$. Besides, the field strength adopts the form
\begin{equation}
F_{\gamma\rho}^{1}=-\frac{1}{4}\theta^{\mu\nu}\left(\{A^0_{\mu},\,\partial_{\nu}F^0_{\gamma\rho}+D_{\nu}F^0_{\gamma\rho}\}-2\{F^0_{\gamma\mu},\,F^0_{\rho\nu}\}\right).\label{eq:field strength-first order}
\end{equation}
We may rewrite this expression in terms of the first order potential $A_{\mu}^{1}$ and the commutative potential $A_{\mu}^{0}$. After some simplifications we obtain
\begin{equation}
\partial_{\gamma}A_{\rho}^{1}-\partial_{\rho}A_{\gamma}^{1}=-\theta^{\mu\nu}\left(A_{\mu}^{0}\partial_{\nu}F_{\gamma\rho}^{0}+\partial_{\mu}A_{\gamma}^{0}\partial_{\nu}A_{\rho}^{0}-F_{\gamma\mu}^{0}F_{\rho\nu}^{0}\right),
\end{equation}
where $F_{ik}^{0}=\partial_{i}A_{k}^{0}-\partial_{k}A_{i}^{0}$ is the field strength tensor at zero order. It follows that
\begin{equation}
F_{\gamma\rho}^{1}=\partial_{\gamma}A_{\rho}^{1}-\partial_{\rho}A_{\gamma}^{1}+\theta^{\mu\nu}\partial_{\mu}A_{\gamma}^{0}\partial_{\nu}A_{\rho}^{0},
\label{fmunu1st}
\end{equation}
an expression that will be useful for later sections.

Based on the previous equations for the first order, two order solutions derived from the SW map by iterative procedure. The following general structure of the solutions can be proposed as~\cite{Ulker:2007fm} 
\begin{eqnarray}
\varLambda_{\lambda}^{n+1} & = & -\frac{1}{4(n+1)}\theta^{\mu\nu}\underset{p+q+r=n}{\sum}\{A_{\mu}^{p},\:\partial_{\nu}\varLambda_{\lambda}^{q}\}_{*r},
\label{eq:parameter n+1 order}
\\[4pt]
A_{\gamma}^{n+1} & = & -\frac{1}{4(n+1)}\theta^{\mu\nu}\underset{p+q+r=n}{\sum}\{A_{\mu}^{p},\:\partial_{\nu}A_{\gamma}^{q}+F_{\nu\gamma}^{q}\}_{*r}.\label{potencialordenn}
\label{eq:gauge potential n+1 order}
\end{eqnarray}
These are recursive relations for the noncommutative fields and by performing similar calculations to the previous ones, we can rewrite the $n$-th order term of the field strength in the form
\begin{eqnarray}
F_{\gamma\rho}^{n+1}&=&-\frac{1}{4(n+1)}\theta^{\mu\nu}\underset{p+q+r=n}{\sum}\left(\{A_{\mu}^{p},\,\partial_{\nu}F_{\gamma\rho}^{q}+(D_{\nu}F_{\gamma\rho})^{q}\}_{*^r} \right.
\nonumber \\[4pt]
&&\left. -2\{F_{\gamma \mu}^{p},\,F_{\rho \nu}^{q}\}_{*^r}\right),
\label{eq:field strength n+1 order}
\end{eqnarray}
where 
\begin{equation}
(D_{\nu}F_{\gamma\rho})^{n} := \partial_{\nu}F_{\gamma\rho}^{n}-i\underset{p+q+r=n}{\sum}[A_{\nu}^{p},\:F_{\gamma\rho}^{q}]{}_{*^r}.
\end{equation}
Here the sum is over all the values of $p$, $q$ and $r$ such that $p+q+r=n$; the subscript $*^r$ in a commutator $[f,g]_{*^r}$ means that we only consider the contributions of the form
\begin{eqnarray}
f(x)*^{r}g(x)&=&\dfrac{1}{r!}\left(\dfrac{i}{2}\right)^{r}\theta^{\mu_{1}\nu_{1}}\cdots\theta^{\mu_{r}\nu_{r}}
\nonumber \\[4pt]
&&\times \partial_{\mu_{1}}\cdots\partial_{\mu_{r}}f(x) \partial_{\nu_{1}}\cdots\partial_{\nu_{r}}g(x).
\end{eqnarray}
Note that $\theta^{\mu\nu}$-tensor appears in the expansions and arise from the mathematical construction; it has the physical interpretation of the noncommutative structure of the space-time.

This is the Seiberg-Witten map \cite{Seiberg:1999vs}: a mechanism to construct noncommutative gauge theories having an explicit dependence on the commutative fields and their derivatives. This work focus on  $U(1)$ Electrodynamics and their NC version. As we shall see in the next section, after reviewing the standard London equation, we obtain the NC version from the theory discussed so far.
%%%%%%%%%%%%%%%%%%%%%%%%%%%%%%%%%%%%%
\section{London equations}
\label{sec3}

Despite the quantum nature of the dissipative process and its relation with the mean free path of electrons, Drude model represents a good approximation for conductivity properties for a large class of materials. In this spirit, we will use the Drude's conductivity for a \emph{perfect} conductor to obtain our NC relation between supercurrent and gauge potential. In other words, the constitutive relation $J=\sigma E$ that is heuristically used to derive the London equation is taken to be valid in our NC version but with a more general linear relation $\hat{J}=\sigma_{eff}\hat{E}$, as we see in the next subsection\footnote{We stress that this procedure is heuristic but effective and coherent with the formal deduction of London relation\cite{London-1935}, see \cite{Annett:730995}.}. We will analyze the phenomenological consequences of NC current, carefully following the role of the SW mappings.

%%%%%%%
Recall that, from Ohm's law and the complex representation of the a.c. currents and fields
\begin{equation}
J^{i}e^{-i\omega t}=\sigma(\omega)E^{i}e^{-i\omega t}.\label{eq:Ohm's law finite frequency}
\end{equation}
Drude's model of conductivity pick up a delta function accounting for dissipationless electrons in a perfect conductor and satisfies the sum rule for $\mathfrak{Re}[\sigma]$\footnote{Drude model is a good approximation if $\omega\mapsto 0$, below the energy gap \cite{Annett:730995}.}
\begin{equation}
\sigma(\omega)=\frac{\pi n_{s}e^{2}}{m_{e}}\delta(\omega)-\frac{n_{s}e^{2}}{i\omega m_{e}}.\label{eq:conductivity}
\end{equation}
Taking the curl of both sides of the Eq.~(\ref{eq:Ohm's law finite frequency})  and using Faraday's law, we obtain a relation between current and potential in the $\omega\mapsto 0$ limit 
\begin{equation}
J^{i}=-\frac{n_{s}e^{2}}{m_{e}}A^{i},\label{eq:London equation}
\end{equation}
where $n_{s}=N_{S}/V$ is the density of superconducting electrons and $A^{i}$ are the components of the magnetic vector potential. Eq. (\ref{eq:London equation}) is the London current for $\sigma$ infinite in perfect conductors and captures another key property that defines a perfect conductor to be a superconductor: the Meissner-Ochsenfeld effect. Indeed, taking the curl in both sides of Eq. (\ref{eq:London equation}) and use Ampere's law (in S.I units)
\begin{equation}
    \nabla\times\left(\nabla\times\textbf{B}\right)=-\nabla^{2}\textbf{B}=-\mu_{}\dfrac{n_{s}e^{2}}{m_{e}}\textbf{B}.\label{LondonETruh}
\end{equation}
The continuity equation is verified by London Equation (\ref{LondonETruh}) provided $\nabla \cdot\textbf{A}=0 $, which is called \emph{London gauge}\footnote{We are considering the static density of superconductor electrons regime, where $\partial_{t}\rho=0$.}. We will analyze the behavior of the London length (or penetration length)\footnote{Recall that we are in the zero temperature limit. Hence, London penetration length is the \emph{bare}, theoretical one. At finite temperature in the Ginzburg-Landau description, behaves like $\lambda (T)=\lambda_{L}\left(1-\left(T/T_{c}\right)^{4}\right)^{-1/2}$ \cite{book:17888}.}
\begin{equation}
    \lambda_{L}^{2} := \dfrac{m_{e}}{\mu_{}n_{s}e^{2}}, \label{londonLength}
\end{equation}
and the NC version of Eq. (\ref{LondonETruh}) under the SW $\theta$-expansions that account for noncommutative gauge potential. To achieve this aim, we consider the gauge potential in complex representation associated with the a.c. current, Eq. (\ref{eq:London equation})
\begin{equation}
A_{\mu}(x_{i},t)=A_{\mu}(x_{i})e^{-i\omega t},\label{eq: Potential A complex representation}
\end{equation}
from which we will get our NC version of London equation. However, we will take a slightly different approach than the previous paragraphs because the NC Maxwell equations have the Moyal product of noncommutative fields that need to be expanded in terms of commutative ones\footnote{Also, the NC gauge parameter Eq. (\ref{eq:parameter n+1 order}) depends on the gauge field but it is not necessary to consider at the moment.}. We recommend the reader take a look at the appendix where a compendium of NC Maxwell equations is presented (\ref{appendix}). As a final note in this paragraph, we stress that Eq. (\ref{eq:London equation}) is the so-called London equation whereas Eq. (\ref{LondonETruh}) is viewed as one of its consequences (using Maxwell equations and vector identities).
%%%%%%%%%%%%%%%%%%%%%%%%%%%%%%%%%%%%%%%%
\subsection{Noncommutative London equations}
An advantageous feature of Eq. (\ref{eq: Potential A complex representation}) is that the $n$-order in $\theta$-expansion of the gauge potential reads
\begin{equation}
    A_{\mu}^{n+1}(x_{i},t)  =A_{\mu}^{n+1}(x_{i})e^{-i(n+2)\omega t}, \label{eq:NC complex gauge potential}
\end{equation}
a result that follows from Eq. (\ref{eq:gauge potential n+1 order}) due to the complex representation of gauge potential. E.g., for first order $A_{\mu}^{1}$ we take $n=0$ from which $A_{\mu}^{1}(x_{i},t)=A_{\mu}^{1}(x_{i})e^{-2i\omega t}$. An induction procedure gives all the rest terms $n\neq 0$ in $\theta$-expansions. Therefore  
\begin{equation}
    A_{\mu}^{n}(x_{i},t)  =A_{\mu}^{n}(x_{i})e^{-i(n+1)\omega t}  \ \ \ \Longrightarrow \ \ \ \begin{cases}
    E_{j}^{n}(x_{i},t) & =E_{j}^{n}(x_{i})e^{-i(n+1)\omega t}\\ \\
    B_{j}^{n}(x_{i},t) & =B_{j}^{n}(x_{i})e^{-i(n+1)\omega t}          
\end{cases}, \label{eq:NC complex gauge potential2}
\end{equation}
for any $n$-order in $\theta$-expansion, where $\ensuremath{B_{i}^{n}=\frac{1}{2}\epsilon_{ijk}F_{n}^{jk}}$ and
$\ensuremath{E_{i}^{n}=F_{n}^{oi}}$ hold, as in commutative case. 

To derive the NC version of the Eqs. (\ref{eq:London equation}) and (\ref{LondonETruh}) we postulate a NC version of the constitutive relation $J=\sigma E$ as
\begin{equation}
\widehat{J}^{i}:=\sigma_{eff}(\omega)\widehat{E}^{i}.
\label{eq:noncommutative Ohm´s law}
\end{equation}
The interpretation of the above equation (which we regard as the NC Ohm's law) is that the effects of NC electric and magnetic fields must affect the physical properties of the material, as the conductivity. For simplicity, our postulate maintain the phenomenological relation between current and vector potential linear in the NC version. As a result of these hypothesis, the \emph{NC conductivity}
\begin{equation}
   \sigma_{eff}(\omega):=\lambda_{eff}^{-2}\pi\delta(\omega)-\frac{\lambda_{eff}^{-2}}{i\omega},\label{eq:conductivity NC}
\end{equation}
acquires an effective penetration length  $\lambda_{eff}^{-2}=\lambda_{eff}^{-2}(A_{i},\theta)$ that accounts for NC effects. As mentioned in the introduction section, we assume \emph{a priori} that NC London penetration length depends on the gauge fields in a non-local way in analogy with Pippard supercurrent\footnote{And in fact, similar to what is called anomalous skin depth effect.} that pickups a $\xi$-parameter and non-local magnetic field dependency.

Under the SW map, the effective penetration length can be expanded as  
\begin{equation}
\lambda_{eff}^{-2}=\lambda_{L_{0}}^{-2}+\theta\lambda_{L_{1}}^{-2} +\theta^{2} \lambda_{L_{2}}^{-2} +\mathcal{O}(\theta^{3}),\label{eq: Penetration length NC}\end{equation}
where $\lambda_{L_{n\neq 0}}^{-2}$ stands for $n$-order $\theta$ perturbation that can be written in terms of commutative (or standard) penetration length $\lambda_{L_{0}}\equiv\lambda_{L}$. For $\theta=0$ the commutative case is recovered. We will compute the Eq. (\ref{eq: Penetration length NC}) up to $\lambda_{L_{2}}$ in the next section.

We procede now to obtain our NC version of Eq. (\ref{eq:London equation}). Analogue to the commutative version described above, we take NC Ampere's law Eq. (\ref{ampere2}), the covariant derivative and the 4-dimensional Levi-Civita tensor in both sides on the NC constitutive relation Eq. (\ref{eq:noncommutative Ohm´s law}), namely
\begin{equation}
\epsilon_{0ijk}D^{j}*\left(D_{\mu}\widehat{F}^{k\mu}\right)=\mu_{}\sigma_{eff}(\omega)\frac{1}{2}\epsilon_{0ijk}D^{0}*\widehat{F}^{jk}+\mu_{}\epsilon_{0ijk}(D^{j}\sigma_{eff}(\omega))\widehat{F}^{0k},\label{eq: Ohm=0000B4s and Faraday  law}
\end{equation}
where the derivative of the conductivity in the second term of r.h.s is due to our postulate $\lambda_{eff}^{-2}=\lambda_{eff}^{-2}(A_{i},\theta)$. Also, we have used  $\widehat{F}^{0i}=\widehat{E}^{i}$ and the NC Faraday's law Eq. (\ref{eq:"Faraday's law"}). The right hand side of the Eq.~(\ref{eq: Ohm=0000B4s and Faraday  law}) can be expanded with the help of Eqs.~(\ref{eq: Faraday's law zero, first and second order}) and~(\ref{eq:NC complex gauge potential}). Up to first order in $\theta$ gives 
\begin{align}
&\mu\sigma_{eff}(\omega)\frac{1}{2}\epsilon_{0ijk}D^{0}*\widehat{F}^{jk}=\nonumber\\
&\mu\sigma_{eff}(\omega)\left[i\omega B_{0}^{i}e^{-i\omega t}+2i\omega B_{1}^{i}e^{-i2\omega t}+\theta^{pq}\left(i\omega A_{p}^{0}-E_{p}^{0}\right)\partial_{q}B_{0}^{i}e^{-i\omega t}+\mathcal{O}(\theta^{2})\right].\label{eq: Faraday complex}
\end{align}
The second term of the right hand side of the Eq.~(\ref{eq: Ohm=0000B4s and Faraday  law}) can be written as
\begin{equation}
\mu_{}\epsilon_{0ijk}(D^{j}\sigma_{eff}(\omega))\widehat{F}^{0k}=\mu_{}\epsilon_{0ijk}D^{j}\sigma_{eff}(\omega)(i\omega A_{k}^{0}e^{-i\omega t}+2i\omega A_{k}^{1}e^{-i2\omega t}+\mathcal{O}(\theta^{2})).\label{eq: Faraday complex part 2}
\end{equation}
At this point, taking $\omega\rightarrow0$ reduces the conductivity, Eq. (\ref{eq:conductivity}), to a constant and above equations tend to
\begin{align}
\frac{\mu}{2}\underset{\omega\rightarrow0}{\lim}\sigma_{eff}(\omega)\epsilon_{0ijk}D^{0}*\widehat{F}^{jk} &=  -\lambda_{eff}^{-2}\left(B_{0}^{i}+2B_{1}^{i}\right)+\mathcal{O}(\theta^{2}),
\label{eq:pendiente1}
\\%[4pt]%
\mu \underset{\omega\rightarrow0}{\lim}\epsilon_{0ijk}(D^{j}\sigma_{eff}(\omega))\widehat{F}^{0k} &=  -\epsilon_{0ijk}(D^{j}\lambda_{eff}^{-2})( A_{k}^{0}+2A_{k}^{1})+\mathcal{O}(\theta^{2}),%\label{potencialordenn}% %DOBLE ETIQUETA%
\label{eq:pendiente2}
\end{align}
where we have used the commutative London current Eq. (\ref{eq:London equation}) to cancel the last term on the right hand side of Eq.~(\ref{eq: Faraday complex}).

Now, let us focus on the left hand side of Eq.~(\ref{eq: Ohm=0000B4s and Faraday  law}). A similar procedure to the case on the right side, allows us to write
\begin{equation}
    \underset{\omega\rightarrow0}{\lim}D_{\mu}*\widehat{F}^{k\mu}=D_{l}*\widehat{F}^{kl}-\theta^{pq}\mu_{}\varepsilon_{0}\partial_{p}A_{0}^{0}(\partial_{q}E_{k}^{0}+\partial_{q}E_{k}^{1})-\theta^{pq}\mu_{}\varepsilon_{0}\partial_{p}A_{0}^{1}\partial_{q}E_{k}^{0}+\mathcal{O}(\theta^{3})=\mu_{}\widehat{J}_{e}^{k},
\end{equation}
and subtituition back on Eq.~(\ref{eq: Ohm=0000B4s and Faraday  law}) togheter with Eqs.~(\ref{eq:pendiente1}) and~(\ref{eq:pendiente2}) gives 
\begin{equation}
\mu_{0} \epsilon_{0ijk}D^{j}*\widehat{J}^{k}=-\lambda_{eff}^{-2}\left(B_{0}^{i}+2B_{1}^{i}\right)-\epsilon_{0ijk}(D^{j}\lambda_{eff}^{-2})( A_{k}^{0}+ 2A_{k}^{1})+\mathcal{O}(\theta^{2}).\label{eq: NC London}
\end{equation}
On the other hand, magnetic fields obtained from field strength up to first
order in $\theta$ are: $B_{0}^{i}=\frac{1}{2}\epsilon_{0ijk}F_{0}^{jk}$,  $B_{1}^{i}=\frac{1}{2}\epsilon_{0ijk}F_{1}^{jk}$ and can be written in terms of the gauge potential 

\begin{equation}
   B_{0}^{i}+2B_{1}^{i}=\epsilon_{0ijk}D^{j}*\left(A_{0}^{k}+2A_{1}^{k}\right), 
\end{equation}
where we have used the Eq.~(\ref{fmunu1st}) to simplify $F_{1}^{jk}$. The London current up to first order in $\theta$ adopts the form
\begin{align}
\widehat{J}^{k} & =-\mu_{}^{-1}\lambda_{eff}^{-2}\left(A_{0}^{k}+2A_{1}^{k}\right)+\mathcal{O}(\theta^{2}),\label{eq: NC London-1}
\end{align}
where $\lambda_{eff}^{-2}$ is the NC version of the London penetration length Eq. (\ref{londonLength}). 

The NC counterpart of Eq. (\ref{LondonETruh}) can be obtained from the NC London current Eq. (\ref{eq: NC London-1}) using NC Ampere's law given in Eq. (\ref{ampere2}) and vector identities. Indeed, focusing on the left hand side of Eq.~(\ref{eq: NC London})
\begin{align}
\epsilon_{0ijk}D^{j}*\left(D_{\mu}*\widehat{F}^{k\mu}\right) & =\left(\nabla\times(\nabla\times[B^{0}+B^{1}])\right)^{i}+\epsilon_{ijk}\epsilon^{klm}\theta^{pq}\partial_{j}\left(\partial_{p}A_{l}^{0}\partial_{q}B_{m}^{0}\right)\nonumber \\
 & -\epsilon_{ijk}\theta^{pq}\partial_{j}\left(\partial_{p}A_{0}^{0}\partial_{q}E_{k}^{0}\right)+\epsilon_{ijk}\epsilon^{klm}\theta^{pq}\partial_{p}A_{j}^{0}\partial_{q}\partial_{l}B_{m}^{0},\label{eq: ED London}
\end{align}
which is an expansion up to first order. Here $\epsilon_{ijk}\epsilon^{klm}\partial_{j}\partial_{l}B_{m}^{n}=\left(\nabla\times(\nabla\times B^{n})\right)^{i}$ and the index $n=0,1$ stand for the perturbation order. Now, using  $\nabla^{2}\hat{B}=\nabla (\nabla\cdot\hat{B} )-\nabla\times (\nabla\times\hat{B} )$ and Eq.~(\ref{eq:Gauss's law zero, first and second order}) we find the London equation for zero and first order
\begin{subequations}\label{londonexp}
\begin{align}
\partial_{l}\partial_{l}B_{i}^{0} &=\lambda_{L}^{-2}B_{i}^{0},\\
\partial_{l}\partial_{l}B_{i}^{1} &=2\lambda_{L}^{-2}B_{i}^{1}+\theta \lambda_{L_{1}}^{-2}B_{i}^{0}+\epsilon_{ijk}\theta\partial_{j}\left(\lambda_{L_{1}}^{-2}\right)A_{k}^{0}-\theta^{pq}\partial_{i}\left(\partial_{p}A_{l}^{0}\partial_{q}B_{l}^{0}\right)
 \nonumber \\ 
 & 
 +\theta^{pq}\epsilon_{ijk}\epsilon^{klm}\left(\partial_{j}(\partial_{p}A_{l}^{0}\partial_{q}B_{m}^{0})+\partial_{p}A_{j}^{0}\partial_{q}\partial_{l}B_{m}^{0}\right).
\end{align}
\end{subequations}
The first equation is the standard London equation (Eq. \ref{LondonETruh}) whereas the second one corresponds to the first order SW $\theta$-expansion. As we will discuss later, we can obtain the London equation to all orders and their solutions using the SW map. In this analysis it is necessary to obtain the NC $\lambda_{L_{i}}$ corrections
of $\lambda_{eff}$ that appear in the above equations; a task that will be made in the next section for a specific system.

As a final note, we stress that we postulate a NC relation between current and A.C electric field with the proportionality constant being the NC conductivity that acquires corrections in $\theta$-expansions, see Eq. (\ref{eq:noncommutative Ohm´s law}). Then, with the knowledge of NC Maxwell equations and vector identities invariant under NC, we derive the relation between NC magnetic field and NC current, order by order in $\theta$-expansions, Eq. (\ref{londonexp}). 
%*****************

\section {Superconductor in noncommutative framework}
\label{secc:8}
\begin{figure}%[H]
\begin{center}
 \includegraphics[width=0.50\textwidth]{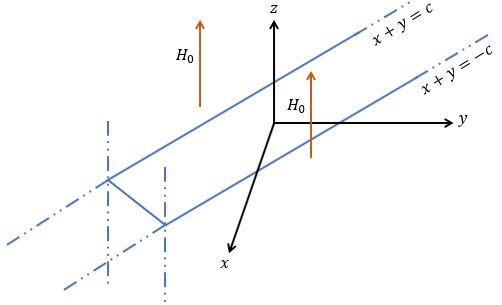}
 % A.png: 427x418 pixel, 72dpi, 15.06x14.75 cm, bb=0 0 427 418
 \caption{The superconductor sample is extended over $y+x=\pm c$ and $z$ directions with thickness $\sqrt{2}c$. A $\pi/4$-rotation will take this configuration to the one whose solution is given in Eq. (\ref{soltink}). The coordinate change between these two arrangements is $\sqrt{2} x'=x+y$, $\sqrt{2} y'=x-y$.}
 \label{fig:0}
 \end{center}
\end{figure}

Now we use the results of the previous section in an specific system. Let be a superconductor slab of thickness $c$ extending infinitely in $z$ and $y$ directions. If we consider a constant external magnetic field $\textbf{H}=H_{0}\hat{z}$ in vacuum space, the microscopic magnetic flux density that solves the London equation (\ref{LondonETruh}) inside the sample reads\footnote{We use the notation $B$ for microscopic magnetic flux density, while we keep $H$ for the external magnetic field. Also, $\mu=\mu_{0}\left(1+\chi_{m}\right)$ such that, in the boundary of the sample, there is no magnetic susceptibility $\chi_{m}$.}
\begin{equation}
    B(x)=\mu H_{0}\dfrac{\cosh(x/\lambda_{L_{0}})}{\cosh(c/2\lambda_{L_{0}})},\label{soltink}
\end{equation}
and shows that $B$ is reduced to a minimum value $\mu H_{0}/\cosh(c/2\lambda)$ at the mid-plane of the slab, i.e., the Meissner effect \cite{book:17888}. In order to perform the SW expansions for London theory, we consider a slightly different configuration, namely, a rotation version of the previous configuration, shown in figure (\ref{fig:0})\footnote{The principal reason for this geometric choice is to perform a more tractable $\theta$-expansions procedure.}.
For the geometric configuration of the figure (\ref{fig:0}), we have been able to obtain a solution for London equation
\begin{equation}
B_{z}(x,y)=\dfrac{\mu H_{0}}{\cosh\left(c/\sqrt{2}\lambda_{L_{0}}\right)}\cosh\left( \dfrac{x+y}{\sqrt{2}\lambda_{L_{0}}}\right) ,  \label{solrotada}
\end{equation}
which reduce to the familiar result Eq. (\ref{soltink}) making a $\pi/4$-rotation. Now, considering $\textbf{B}=\nabla\times\textbf{A}$, the non-zero components of vector potential are 
\begin{equation}
 A_{x}\equiv A_{1}^{}  = - \dfrac{\sqrt{2}\lambda_{L_{0}} \mu H_{0}}{2\cosh\left(c/\sqrt{2}\lambda_{L_{0}}\right)}\sinh\left( \dfrac{x+y}{\sqrt{2}\lambda_{L_{0}}}\right)=-A_{y}\equiv -A_{2}.
\label{eq:Potential A0}
\end{equation}
With the help of Eqs.~(\ref{solrotada}), (\ref{eq:Potential A0}) and exploiting the benevolence of hyperbolic functions under the derivative operator, we now proceed to achieve our SW maps that determine the analytic expressions of the gauge potentials $A_{\mu}$ to all orders in $\theta$. If the sample is sited in the $x y$-plane we need fix the values  $\theta^{12}=-\theta^{21}=\theta$ and the remaining components are taken to be zero.
\subsection{Solution of the noncommutative London\textquoteright s equation}
%\label{secc:8}

Exploring the gauge potential order by order explicitly, some regularities appear. After a considerable but straightforward procedure, the spatial components of noncommutative gauge potential $\widehat{A}_{\mu}$ up to third order on perturbation theory read %{\color{red}YA NO HAY VARIAS LAMBDAS $\lambda_{li}$}
\begin{eqnarray}
A_{k}^{1} & = & \theta A_{k}^{0}B_{z}^{0},\nonumber \\[4pt]
A_{k}^{2} & = & \frac{1}{2}\theta(A_{k}^{1}B_{z}^{0}+A_{k}^{0}B_{z}^{1})=\theta^{2} A_{k}^{0}[(A_{k}^{0})^{2}\lambda_{L_{0}}^{-2}+(B_{z}^{0})^{2}],\nonumber \\[4pt]
A_{k}^{3} & = & \frac{1}{3}\theta(A_{k}^{0}B_{z}^{0}+A_{k}^{1}B_{z}^{1}+A_{k}^{0}B_{z}^{2}),\label{eq:S-WSolutions for A}
\end{eqnarray}
and so on. Here, $A_{k}^{0}$ is the zero-order gauge potential (that is, the commutative one). %{\color{red} and $\epsilon_{ij}$ stands for $2$-dimensional Levi-Civita tensor. 
An induction argument allows us to write  for all orders
  \begin{equation}
  A_{x}^{n}=-A_{y}^{n}. \label{genform1}
 \end{equation}
%From the last equation, we can read off the coefficients $C_ {n>0}$, $C_{1}=2\theta/\lambda_{L_{i}},$ $C_{2}=6\theta^{2}/\lambda_{L_{i}}^{2}$ and $C_{3}=64\theta^{3}/3\lambda_{L_{i}}^{3}$.} 

It turns out that Eq. (\ref{genform1}) follows from the general form of SW mappings, Eq. (\ref{potencialordenn})
\begin{equation}
A_{k}^{n+1}=\frac{\theta}{n+1}\underset{p+q=n}{\sum}A_{k}^{q}B_{z}^{p},
\end{equation}
that satisfies 
$\partial_{i}A_{j}^{n}=\epsilon_{ijk}\frac{1}{2}B_{k}^{n}.$

\begin{figure}%[ht]
%\begin{subfigure}{.5\textwidth}
  \centering
  % include first image
  \includegraphics[width=.95\linewidth]{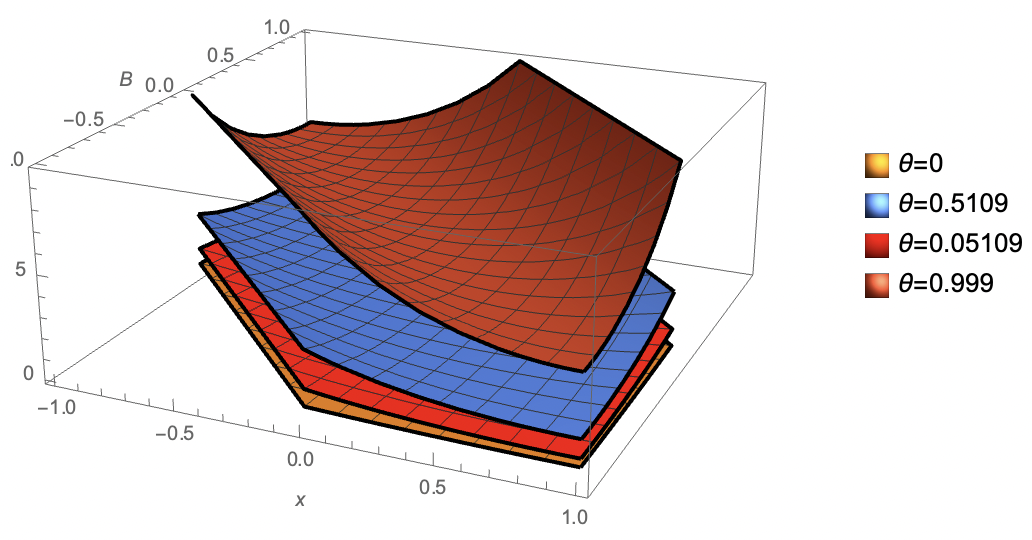}  
  \caption{Behavior of the ${B}$-field in two dimensions.}
  \label{fig:sub-first}
%\end{subfigure}
\end{figure}

\begin{figure}
%\begin{subfigure}{.5\textwidth}
  \centering
  % include second image
  \includegraphics[width=.85\linewidth]{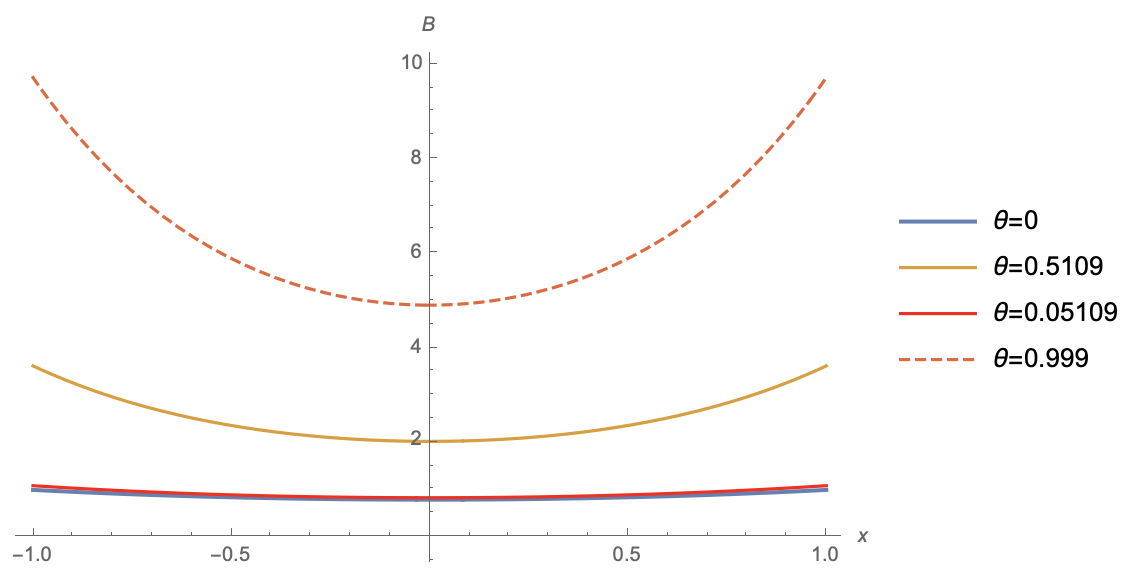}  
  \caption{
   Cross section of the ${B}$-field in $\hat{z}$-direction.}
  \label{fig:sub-second}
%\end{subfigure}
\label{fig:fig}
\end{figure}

\begin{figure}
\begin{center}
 \includegraphics[width=1.0\textwidth]{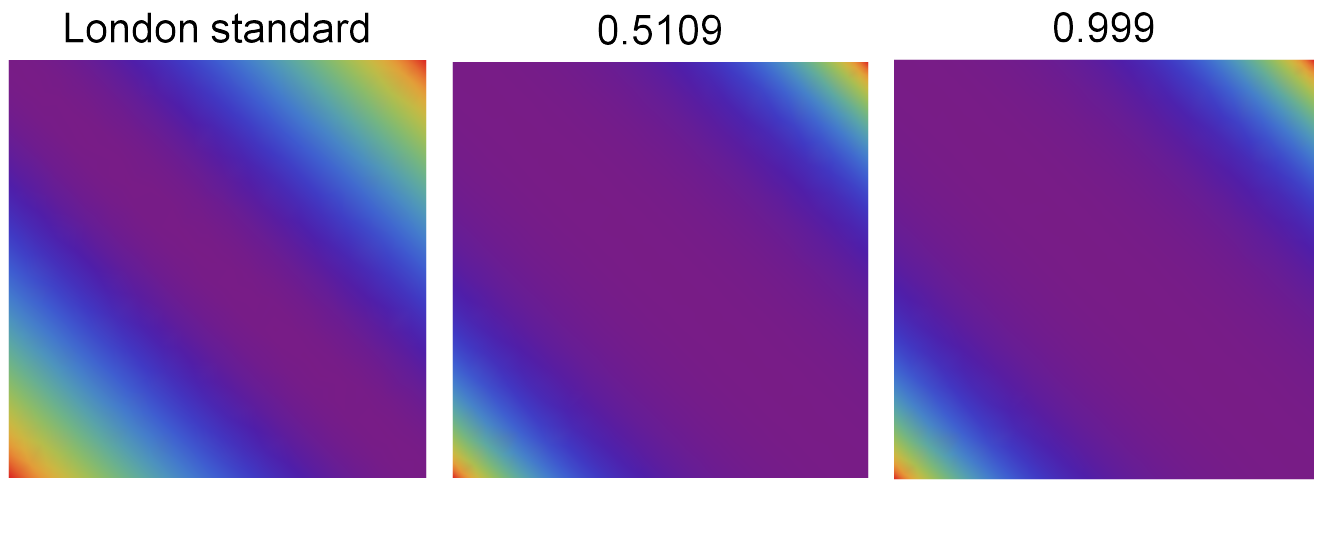}
 % A.png: 427x418 pixel, 72dpi, 15.06x14.75 cm, bb=0 0 427 418
 \caption{
Density plot comparison of the field $B/\mu H_{0}$ between the commutative (left) case and NC case, considering the geometry of figure (\ref{fig:0}) for values $\lambda_{L_{0}}=1$ and superconductor thickness $\sqrt{10}$. In the purple colored region, at the core of superconducting sample, the field tends to zero (Meissner effect). The two NC cases correspond to $\theta=0.5109$ (middle) and $\theta=0.999$ (right). As can be seen, NC causes faster fall-offs of $B$-field, from the boundary to the interior.}
\label{fig:1.1}
\end{center}
\end{figure}

%With these considerations, we are able to demonstrate that the following expression holds, for all orders 
%\begin{equation}
%A_{x}^{n+1}=-A_{y}^{n+1}. \label{Gauge_potential_relation}
%\end{equation}
 Hence, using  $\ensuremath{\widehat{B}_{i}=\frac{1}{2}\epsilon_{ijk}\widehat{F}^{jk}}$, the NC $B$-fields induced by the SW mapping up to second order reads
\begin{alignat}{1}
B_{z}^{0} & =\dfrac{\mu_{}H_{0}}{\cosh\left(c/\sqrt{2}\lambda\right)} \cdot \cosh\left( \dfrac{x+y}{\sqrt{2}\lambda}\right) ,\nonumber \\
B_{z}^{1} & =\theta [2(A_{x}^{0})^{2}\lambda_{L_{0}}^{-2}+(B_{z}^{0})^{2}],\nonumber \\
B_{z}^{2} & =\theta^{2} B_{z}^{0}[7(A_{x}^{0})^{2}\lambda_{L_{0}}^{-2}+(B_{z}^{0})^{2}].\label{Magnetic_field NC}
\end{alignat}
Despite the \emph{non-self-evident} nature of NC fields, the SW map allows us to write the  noncommutative $B$-field up to second order in $\theta$ as  $\hat{B}=B_{z}^{0}+B_{z}^{1}+B_{z}^{2}+\mathcal{O}(\theta^{3})$. In terms of the commutative fields, this expansion is
\begin{equation}
\hat{B} =B_{z}^{0}+ 2\theta(A_{x}^{0})^{2}\lambda_{L_{0}}^{-2}+\theta(B_{z}^{0})^{2}+7\theta^{2} B_{z}^{0}(A_{x}^{0})^{2}\lambda_{L_{0}}^{-2}+\theta^{2}(B_{z}^{0})^{3}+\mathcal{O}(\theta^{3})\cdots. \label{eq: noncamp}
\end{equation}
This equation emerges directly from SW map. However, we need to verify its consistency with the London equations; a task that will be developed in the next section. As we shall see, this analysis also allows us to write down an expression of London penetration length with NC corrections. Finally, for some values of the noncommutative parameter in Eq.~(\ref{eq: noncamp}), the figures (\ref{fig:sub-first}) and (\ref{fig:sub-second}) show the behavior of the NC $B$-field relative to the commutative one. We see that the penetration rate of the field is greater than the commutative field. Figure (\ref{fig:1.1}) shows a density plot perspective.
%%%%%%%%%%%%%%%%%%%%%%%%%%%%%%%%%%%%%%%%%%%%%%%%%%%%%%%%%%%%%%%%%%%%%%%%%%
\subsection{Noncommutative penetration length}
 From the NC London equation (\ref{eq: NC London-1}), 
 %$\hat{J}_{i}=-\mu_{}^{-1}\lambda_{eff}^{-2}\left(A_{0}^{k}+2A_{1}^{k}\right)+\mathcal{O}(\theta^{2})$, we can write the NC solution of Eq.~(\ref{solrotada})
namely
\begin{equation}
\hat{B}\equiv \dfrac{\mu H_{0}}{\cosh\left(c/\sqrt{2}\lambda_{L_{eff}}\right)} \cdot \cosh\left( \dfrac{x+y}{\sqrt{2}\lambda_{L_{eff}}}\right).\label{eq: NC depth}
\end{equation}
The effective penetration length is explicitly given by (using the solution of London equation, Eq.~(\ref{eq: NC depth}) and Eq.~(\ref{eq: noncamp}))
\begin{equation}
\frac{1}{\lambda_{L_{eff}}}=\frac{1}{\lambda_{L}}-\frac{\sqrt{2}}{2(x+y)}\ln\left({1+ \frac{2{\cosh\left(c/\sqrt{2}\lambda_{L_{0}}\right)}}{\mu H_{0}} e^{\frac{x+y}{\sqrt{2}\lambda_{L_{0}}}}(B_{z}^{1}+B_{z}^{2}+\mathcal{O}(\theta^{3}))+\cdots}\right).
\label{loneff}
\end{equation}
There is no cast of doubt that if $\theta\rightarrow0$ then $\lambda_{L_{eff}}\rightarrow\lambda_{L}$, as can be seen from the expressions given in Eq. (\ref{Magnetic_field NC}). In other words, the classical case is consistently obtained within these limits. This expression depends on space coordinates due to the fact that our effective penetration length depends on the potential $A_{i}$ and, when we perform the SW map to obtain the $\theta$-expansions in terms of commutative fields, this dependency arises. We argue that NC induces this coordinate dependency due to the intrinsically anisotropic nature of NC electrodynamics \cite{Adorno:2011wj}. More remarks about this result  can be read in the conclusions section (\ref{conclusiones}). 


\begin{figure}
\begin{center}
 \includegraphics[width=0.70\textwidth]{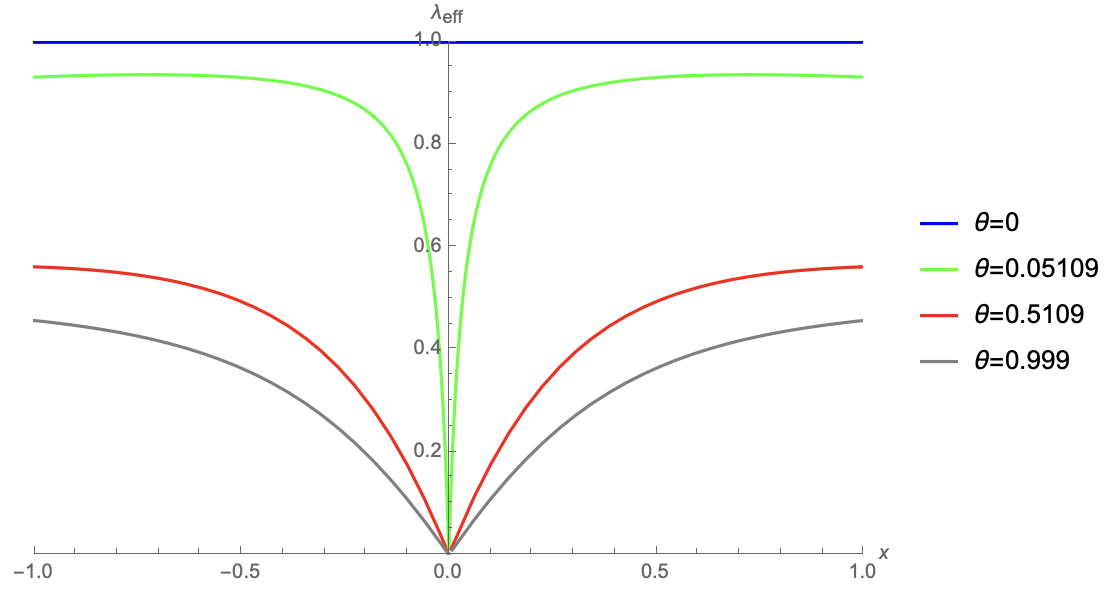}
 % A.png: 427x418 pixel, 72dpi, 15.06x14.75 cm, bb=0 0 427 418
 \caption{\begin{small}Penetration length of Eq. (\ref{loneff}) up to second-order taking several values of the noncommutative parameter $\theta$. Faster fall-offs of the $B$-field are consequence of the increasing penetration length due to $\theta$-expansion. \end{small}}
\label{fig:2}
\end{center}
\end{figure}

Examining the Eq.~(\ref{eq:gauge potential n+1 order}) and using the %recurrence
relations of gauge potential Eq. (\ref{genform1}) into  Eq.~(\ref{eq:field strength n+1 order}), the $B$-field obeys
\begin{equation}
\epsilon_{ijk}B_{n}^{k}=F_{ij}^{n}=\partial_{i}A_{j}^{n}-\partial_{j}A_{i}^{n},\label{eq:Magnetic field n-order}
\end{equation}
where $n=0,1,\dots, m$ stands for the perturbation order. Using these relations and following the same approach of section~(\ref{sec3}), the noncommutative corrections of the London equation take the form
%all orders in $\theta$ takes the form
\begin{align}
\left(\partial_{xx}^{2}+\partial_{yy}^{2}\right)B_{z}^{0} & =\lambda_{L_{0}}^{-2}B_{z}^{0},\nonumber \\
\left(\partial_{xx}^{2}+\partial_{yy}^{2}\right)B_{z}^{1} & =2\lambda_{L_{0}}^{-2}B_{z}^{1}+\theta\lambda_{L_{1}}^{-2}B_{i}^{0}+\epsilon_{ijk}\theta\partial_{j}\left(\lambda_{L_{1}}^{-2}\right)A_{k}^{0},\nonumber \\
\left(\partial_{xx}^{2}+\partial_{yy}^{2}\right)B_{z}^{2} & =3\lambda_{L_{0}}^{-2}B_{z}^{2}+2\theta\lambda_{L_{1}}^{-2}B_{z}^{1}+\theta^{2}\lambda_{L_{2}}^{-2}B_{z}^{0}+2\epsilon_{ijk}\theta\partial_{j}\left(\lambda_{L_{1}}^{-2}\right)A_{k}^{1}+\epsilon_{ijk}\theta\partial_{j}\left(\lambda_{L_{2}}^{-2}\right)A_{k}^{0}.\label{eq:NCLondo ED}
\end{align}
As defined in previous sections, $\lambda_{L_{0}}^{-2}$ ($\lambda_{L_{n}}^{-2}$) is the commutative (noncommutative) London penetration length, respectively.

In order to obtain the NC contributions $\lambda_{L_{1}}^{-2}$ and $\lambda_{L_{2}}^{-2}$, we use the expression for the $B$-field Eq.~(\ref{Magnetic_field NC}) into Eq.~(\ref{eq:NCLondo ED})
\begin{equation}
    \begin{array}{rcl}
         \lambda_{L_{1}}^{-2} &=&2\lambda_{L_{0}}^{-2}B_{z}^{0},  \\ \\
         \lambda_{L_{2}}^{-2} &=& \lambda_{L_{0}}^{-2}\left(3(B_{z}^{0})^{2}+4\lambda_{L_{0}}^{-2}(A_{1}^{0})^{2}\right),
    \end{array}
\end{equation}
then, from Eqs.~(\ref{eq:conductivity NC}) and~(\ref{eq: Penetration length NC}), the NC effects of conductivity $\widehat\sigma_{eff}(\omega)$ is given by
\begin{equation}
\lambda_{L,eff}^{-2}=\lambda_{L_{0}}^{-2}(1+2\theta B_{z}^{0}+3\theta^{2} (B_{z}^{0})^{2}+4\theta^{2} (A_{1}^{0})^{2}+\mathcal{O}(\theta^{3})) \label{maineq2}.
\end{equation}

This equation is our main result. Considering our postulate NC Ohm's law Eq. (\ref{eq:noncommutative Ohm´s law}), and performing the noncommutative SW map on gauge fields, we obtain an effective London penetration length up to second order in $\theta$-parameter. Also, $\lambda_{L, eff}$ is expressed in terms of the density of superconducting electrons $n_{s}$ that forms the supercurrent, Eq. (\ref{londonLength}) but, due to the lack of quantum framework (a Ginzburg-Landau model), we cannot argue, up to this point, that this quantity is been affected by NC\footnote{Moreover, in type-I superconductors that verifies London equation, the superconductor electronic density $n_{s}$ is constant.}. In other words, Eq. (\ref{maineq2}) can be see as a phenomenological implication of NC theory. In figure (\ref{fig:2}), we show the effective London length $\lambda_{Leff}$ as a function of distance from the boundary of the sample. More details of this result and its possible extensions can be read in the conclusion section (\ref{conclusiones}). 

\begin{figure}
\begin{center}
 \includegraphics[width=0.5\textwidth]{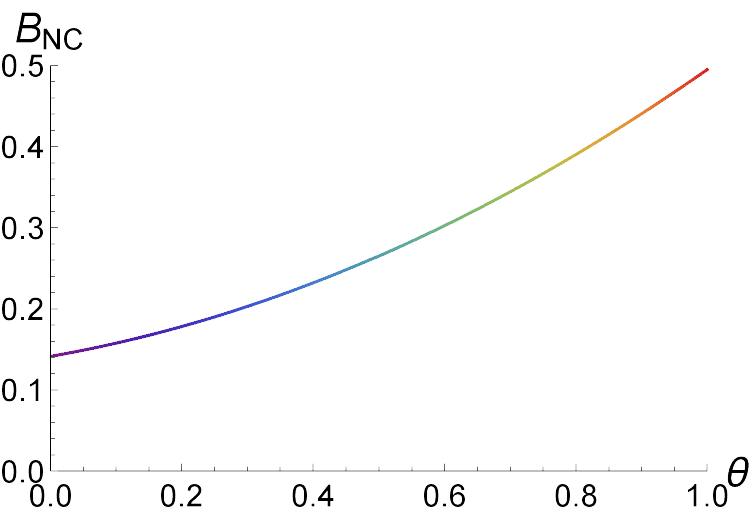}
 % A.png: 427x418 pixel, 72dpi, 15.06x14.75 cm, bb=0 0 427 418
 \caption{Qualitative behavior of the macroscopic noncommutative $\mathcal{B}_{NC}$-field for different values of $\theta\in \left[0,1\right]$. In this plot, we take $\mu H_{0}=1=\lambda_{L_{0}}$, $c/\lambda_{L_{0}}=10\sqrt{2}$. When $\theta=0$, the commutative $B$-field is recovered.}
\label{fig:1.4}
\end{center}
\end{figure}

To end this section, a macroscopic flux density $\mathcal{B}$ is obtained averaging over the sample thickness $\sqrt{2}c$ \cite{book:17888}. For the commutative standard case, Eq.~(\ref{solrotada})
\begin{equation}
\mathcal{B}\equiv\Bar{B}=\mu \textbf{H}=\mu_{0}\left(1+\chi_{m}\right)\textbf{H}=\mu H_{0} \left( \dfrac{\sqrt{2}\lambda_{L_{0}} }{c} \right) \tanh\left( \dfrac{c}{\sqrt{2}\lambda_{L_{0}}} \right),
\end{equation}
where $\textbf{H}$ and $\textbf{M}$ obey the linear constitutive relation $\textbf{M}=\chi_{m}\textbf{H}$, being $\textbf{M}$ the magnetization. In the NC case, we need to integrate order by order the Eq.~(\ref{eq: noncamp}), namely
\begin{align}
&\mathcal{B}_{NC}\equiv\Bar{B}_{NC} =\nonumber \\
&\mathcal{B} \left[1+\theta \mu H_{0} +\theta^2\dfrac{\mu^{2} H_{0}^2}{4} \left( 3 \cosh \left({\sqrt{2} c}/{\lambda_{L_{0}}}\right)+1\right) \text{sech}^2\left({c}/{\sqrt{2} \lambda_{L_{0}}}\right)+\mathcal{O}(\theta^{3})
\right].
\end{align}
In the figure (\ref{fig:1.4}) we show the macroscopic magnetic flux average as a funtion of $\theta$.
%************************

\section{Flux quantization in London theory }%\label{sec:DQC-in-theWu-Yang approach} 
\label{secc:4}

The order parameter of the condensate $\Psi(\textbf{r})$ should be a single-valued
function, leading to a quantization of the magnetic flux. In this section we prove that the NC effects controlled by SW mappings verifies this statement. 

Let us consider a hole in a superconductor bulk with the trapped magnetic
flux 
\[
\varPhi_{B}=\int d\textbf{r}\cdot\textbf{B},
\]
where the surface integral is taken over the cross section, which includes the hole. The $B$-field does not penetrate into the bulk deeper than $\lambda_{L}$. Hence, we can find a contour C surrounding the hole, along which the field and the current are zero. Normalizing the condensate wave function
as $\Psi(\textbf{r})=\sqrt{n_{s}}\exp(i\lambda)$ and taking $\textbf{j}=0$, we can express the vector potential along the contour as\footnote{We apologies for the use of $\lambda$ as the phase of order parameter, hoping the reader don't be confused with the penetration length parameter. The reason of this choice is due to that $\lambda$ ($\Lambda$) is the gauge parameter in commutative $U(1)$ (noncommutative $U_{*}(1)$) gauge theory, as stated in the section (\ref{sec2}).}
\begin{equation}
A_{i}(\textbf{r})=\partial_{i}\lambda.\label{eq: flux gauge}
\end{equation}

Then the magnetic flux becomes 
\[
\varPhi_{B}=\oint_{C}d\textbf{r}\cdot\textbf{A}(\textbf{r})=\frac{\delta\lambda}{e^{*}},
\]

where we have put $e^{*}$ explicitly. Here $\delta\lambda$ is a change of phase in the round trip along
the contour. The wave function is single-valued if $\delta\lambda=2\pi n$
where $n=0,1,2,\ldots$ Hence, the flux is quantized ($\varPhi_{B}=\varPhi_{0}n$), and the quantum flux in units of $2\pi\hbar c/e^{*}$ is given by 
\[
\varPhi_{0}(n=1)=\frac{2\pi\hbar c}{e^{*}}=2.07\times10^{-7}\textrm{G\ c\ensuremath{m^{2}}},
\]
for $e^{*}=2e$ it is reduce to the one observed experimentally in \cite{PhysRev.79.845}. 

We briefly discussed the basic idea towards the generalization of the above result to noncommutative case in one coordinate direction, say $x$. Firstly, the noncommutative gauge potentials $\ensuremath{\hat{A}_{\mu}^{\prime}(x)}$ and $\hat A_{\mu}(x)$ are related by the gauge transformation, Eq.~(\ref{eq:gauge transformation-1}).
Secondly, the noncommutative wave function $\hat{\varPsi}$ change under noncommutative gauge transformation as \begin{equation}
\ensuremath{\hat{\varPsi}^{\prime}(x)}\rightarrow\hat{\varPsi}(x)=U*\hat{\varPsi}(x),\label{eq: NC wave function}
\end{equation}
where $U$ is an element of the noncommutative group $U_{*}(1)=e^{i\hat{\varLambda}(x)}$ according to the SW map. Finally, the noncommutative wave function $\hat{\varPsi}$ should be  single-valued.

The properties discussed above are analogs to the commutative ones. If we assume that the noncommutative wave function is given by $\ensuremath{\hat{\varPsi}^{\prime}(x)}=e^{i\hat{\varLambda}(x)}*\hat{\varPsi}(x)$ and considering that it makes a round trip along contour $C$ then, it generates a change of phase $\delta\hat{\varLambda}$ given by the round trip along the contour. Therefore, the noncommutative wave function is single-valued if $\delta\hat{\varLambda}=2\pi n$ where $n=0,1,2,\ldots$, is the winding number.

Now, we will follow a perturbative treatment where the noncommutative  gauge parameter admits the 
series development given in the Eq.~(\ref{eq:gaugepar}). A change of phase reads
\begin{equation}
\delta\widehat{\Lambda}=2\pi n+\sum_{m=1}^{\infty}\delta\Lambda^{m}.\label{eq: Fase NC}
\end{equation}
It follows that if the flux quantization  is preserved then the noncommutative contributions to the gauge parameter should vanish.
%**************************
\subsection{The noncommutative parameter}
\label{sub:The-noncommutative-parameter}
%$\widehat{\Lambda}=\widehat{\Lambda}\left(\lambda,\,A; \theta \right)$}

Having arrived to a general ansatz for the noncommutative expansions of the gauge potentials in the superconductor, we now proceed to discuss the flux quantization. The main point to analyze is if the $\theta$-corrections applied to the standard gauge parameter $\lambda$ can give vanished values when the gauge potentials are obtained from the SW map. Indeed, from Eq.~(\ref{eq:parameter first order}), the noncommmutative parameter $\widehat{\Lambda}$ to first order is
\begin{eqnarray}
\Lambda^{1}=-\dfrac{1}{2}\theta^{kl}A^0_{k}\partial_{l}\lambda.
\end{eqnarray}
The gauge parameter plays an important role in the quantization of the magnetic flux. Using Eq.~(\ref{eq: flux gauge})  into the noncommutative parameter $\Lambda^{1}$ to first order, we get
\begin{equation}
\Lambda^{1}=-\dfrac{\theta}{2}\epsilon^{kl}A_{k}^{0}A_{l}^{0}=0,\label{eq:Lamda first order=00003D00003D 0}
\end{equation}
where we have assumed $\theta^{12}=-\theta^{21}=\theta$ as the only non-vanishing components and, in consequence, $\theta^{kl}=\theta\epsilon^{kl}$ with $k, l=1,2,3$. Therefore, using the SW map the Flux quantization of the $B$-field is preserved to first order. For second order $\Lambda^{2}$-case we need take into account the general expression Eq. (\ref{eq:parameter n+1 order}). For $n=1$ turns out to be 
\begin{eqnarray}
\Lambda^{2} & = & -\dfrac{1}{4}\theta^{kl} \left( \left\{ A_{k}^{1},\,\partial_{l}\lambda\} + \{A^0_{k},\,\partial_{l}\Lambda^{1} \right\} \right).
\end{eqnarray}
The second term of this equation is zero, since we have already shown that $\Lambda^{1}=0$. Therefore, we only need to calculate $\theta^{kl} \{A_{k}^{1},\,\partial_{l}\lambda \}$. A similar procedure to the calculation of $\Lambda^{1}$ together with the relation given in Eq.~(\ref{genform1}) shows that the gauge parameter to second order also vanishes, i.e. $\Lambda^{2}=0$. 

To discuss the general $n$-order case, first we recall the fact that the noncommutative corrections to the gauge parameter have the general form
\begin{eqnarray}
\Lambda^{n+1} & = & -\dfrac{1}{4(n+1)}\theta^{kl}\underset{_{p+q+r=n}}{\sum}\{A_{k}^{p},\,\partial_{l}\Lambda^{q}\}_{*r}.
\end{eqnarray} 
It is straightforward to see that the $\mathcal{O}(\theta^{r})$ contribution for the anticommutator $\{A_{k}^{p},\,\partial_{l}\Lambda^{q}\}_{*r}$, for $r$ an even number vanishes. Therefore, we can write
\begin{eqnarray}
\Lambda^{n+1} 
& = & -\dfrac{1}{4(n+1)}\theta^{\mu_{1}\nu_{1}}\left(\{A_{\mu_{1}}^{n},\,\partial_{\nu_{1}}\lambda\}+\{A_{\mu_{1}}^{n-1},\,\partial_{\nu_{1}}\Lambda^{1} \} \right.
\nonumber \\[4pt]
&&+...+\{A_{\mu_{1}},\,\partial_{\nu_{1}}\Lambda^{n}\} +\{A_{\mu_{1}}^{n-1},\,\partial_{\nu_{1}}\lambda\}_{*1}
\nonumber \\[4pt]
 &  & +...+\{A_{k}^{0},\,\partial_{\nu_{1}}\Lambda^{n-1}\}_{*1}
 \nonumber \\[4pt]
 &  & \qquad\hfill\:\vdots
 \nonumber \\[4pt]
 &  & \:+\{A_{\mu_{1}}^{1},\,\partial_{\nu_{1}}\lambda\}_{*n-1}+\{A_{k}^{0},\,\partial_{\nu_{1}}\Lambda^{1}\}_{*n-1}
 \nonumber \\[4pt]
 &  & \left.+\{A_{\mu_{1}}^{0},\,\partial_{\nu_{1}}\lambda\}_{*n}\right).
\end{eqnarray} 
We have already seen that the noncommutative corrections $\Lambda^{1}$ and $\Lambda^{2}$ vanish. Let us now assume that this holds up to the $n$-th order, i.e. $\Lambda^{n}=0$. We would like to show that this assumption implies that the expression 
\begin{eqnarray}
\Lambda^{n+1} &=&  -\dfrac{1}{4(n+1)}\theta^{\mu_{1}\nu_{1}}\left(\{A_{\mu_{1}}^{n},\,\partial_{\nu_{1}}\lambda\} \right.
\nonumber \\[4pt]
&& +\sum_{s=1}^n\left. \{A_{\mu_{1}}^{n-s},\,\partial_{\nu_{1}}\lambda\}_{*s}\right)
\label{eq:Lamda (N+1); Lamnda N=00003D0}
\end{eqnarray}
for the $(n+1)$-th order also vanishes.

From Eq.~(\ref{genform1}), we obtain for the first term of Eq.~(\ref{eq:Lamda (N+1); Lamnda N=00003D0}) that $\ensuremath{\theta^{kl}\{A_{k}^{n},\,\partial_{l}\lambda\}}=2\theta^{kl}A_{k}^{n}A_{l}^{0}=2\theta^{}(A_{1}^{n}A_{2}^{0}-A_{2}^{n}A_{1}^{0})=0$; where we have used $A_{i}^{0}=\partial_{i}\lambda$. 
For the second term we need compute the general expression
\begin{align*}
\ensuremath{\theta^{kl}\{A_{k}^{n-s},\,\partial_{l}\lambda\}_{*s}} & =2\theta^{kl}A_{k}^{n-s}*^{s}A_{l}^{0}
.\end{align*}
It is easy to check that $\theta^{kl}A_{k}^{n-s}*^{s}A_{l}^{0}=\theta^{}(A_{1}^{n-s}*^{s}A_{2}^{0}-A_{2}^{n-s}*^{s}A_{1}^{0})=0$ if we use that $A_{1}^{n}=-A_{2}^{n}$ for all orders. In consequence $\Lambda^{n+1}=0$ for all $n$ and therefore $\widehat{\Lambda} = \lambda$. Then, the Flux quantization is preserved under noncommutative corrections coming from the gauge potentials obtained via the SW map.
%%%%%%%%%%%%%%%%%%%%%%%%%%%%%%%%%%%%%%%%%%%%%%%%%%%%%%%%%%%%%%%%%%%%
%\newpage
\section{Discussion and concluding remarks}\label{conclusiones}

In this work, we study the effects of noncommutative (NC) electrodynamics in London-Meissner superconductivity. Using the Seiberg-Witten map, the $\theta$-expansion allowed us to explore a few consequences of NC in London equation. We find that the NC framework and flux quantization condition for the magnetic flux density are compatible with each other throughout the consistency of gauge symmetry and we drive explicit expressions for the Maxwell A-potentials, using London gauge, that satisfy the external NC $B$-field surrounding the superconductor. 

Analyzing the structure of London equations and, by demanding the Seiberg-Witten map to stay well-posed, we find that the noncommutative effects can be incorporated in an \emph{effective} London penetration length; justified by the theoretical development of Pippard on non-local superconductivity. In this work, the effective London penetration length acquires corrections in Seiberg-Witten expansions in terms of the $\theta$-parameter, related with the Pippard product $a\xi_{0}$. As consequence, the magnetic flux inside the superconductor has faster fall-offs relative to the London equation.

We start by considering that Drude's model of conductivity for a perfect conductor is taken to be valid and we promoted it to an NC version.  In the latter case, one is tempted to argue that conductivity is explicitly affected by effective London penetration length and implicitly by the density of Copper pair electrons $\Psi\sim 2ne$. Nevertheless, at the stage of our work, it is not clear a priory (and by the lack of noncommutative BCS theory) that NC effects can change the density of Cooper pairs. Our effective London penetration length is interpreted in this work as a phenomenological consequence of NC expansions.

More phenomenological interpretations of our results can give perspective into the microscopic interactions, keeping in mind that our noncommutative version of London penetration length is considered at zero temperature where the quantum fluctuations are expected to be important relative to thermal ones. A great discussion on specialized literature about the ground state of general condensed matter systems reveals that entanglement is a truly fundamental quantity that controls the interactions between Cooper electrons. An open question that we are interested to explore is if there is an underlying scheme in which we can explore the effects of noncommutative geometry on superconductors in the entangling regime, usually with strong coupling. A very interesting recent work shows that noncommutative geometry induces entangling effects in the anisotropic harmonic oscillator\footnote{See \cite{Muhuri:2020did}.} and other study reveal that entanglement can also arise in NC harmonic oscillator system, with mathematical structure reminiscent of the Unruh effect by using the Landau problem in the presence of a harmonic interaction~\cite{Pal:2020jvq}. NC Electrodynamics is intrinsically anisotropic, therefore, a full Ginzburg-Landau approach with NC effects is feasible.

In this context, generalizations of London current that consider non-spherical geometry of Fermi surface (and also beyond the Drude model) are in terms of anisotropic dispersion of superconducting electrons\footnote{The consensus is that anisotropy is related with the lattice structure that breaks invariance under translations.}, where the London penetration length acquires anisotropy. In this work, anisotropic London penetration length (coordinate dependency) its a consequence of keep NC electrodynamics well posed so, in absence of more deep interpretations, we argue that our NC mathematical setup is capable to explore these issues in a more fundamental quantum Fermi theory.

Another interesting aspect to investigate with NC formalism is, if entanglement close to the Planck scale, can be resolved in the context of noncommutative geometry of space-time and how this behaves on higher distance scales, e.g. atomic scales, where the role of entanglement is not fully understood and the effects of noncommutative have no clear phenomenological interpretation. The $\theta^{\mu\nu}$ tensor can be viewed as an area in which the effects of noncommutative geometry are expected to be important; analogous to Planck constant in ordinary quantum mechanics. Hence, in this work, we assert that despite that NC electrodynamics operates at ultra UV scales, noncommutative geometry has formal and mathematical grounds\footnote{see \cite{PhysRevD.64.067901} for explore the effects on the spectrum by noncommutative quantum mechanics.} under which we can explore these ideas. Other works emphasize the role of $\theta$-scalar as a sort of background primordial Electromagnetic field\footnote{See \cite{Berrino:2002ss}.} therefore, it could be interesting to drive our $\theta$-expansions in gauge fields starting from the complete Ginzburg-Landau Field Theory,  in order to contribute to this unproven claim. 

\subsection*{Remarks about SW map in more general theories}

A natural generalization of this work is in the context of a \emph{noncommutative Ginzburg-Landau} theory (NCGL), and must start with a free energy functional. In the case of London limit, for which the order parameter is constant up to an arbitrary phase $\Psi=\sqrt{n}_{s}\exp(i\phi(r))$, the free energy functional can be written as
\begin{equation}\label{LEnergy}
\hat{F}_{s}=\hat{F}_{s}^{(0)}+\hat{\rho}_{s}\int d^{3}r\left(\nabla\hat{\phi}+\dfrac{2e}{\hbar}\hat{\textbf{A}}\right)*\left(\nabla\hat{\phi}+\dfrac{2e}{\hbar}\hat{\textbf{A}}\right),
\end{equation}
being the NC \emph{superfluid stiffness} $\hat{\rho}_{s}=\left(\hbar/2m^{*}\right)\hat{\bar{\Psi}}*\hat{\Psi}$  \cite{Annett:730995}. At this stage, the NC nature of supercurrent is captured by Moyal-Groenewold product. Then, it follows the NC supercurrent equation 
\begin{equation}\label{ourJ1}
\hat{\textbf{J}}_{s}=-\dfrac{\delta F}{\delta \hat{\textbf{A}}}=-\dfrac{(2e)^{2}}{2m^{2}}\hat{\bar{\Psi}}*\hat{\Psi}\left(\hat{\textbf{A}}\right),
\end{equation}
%{\color{green} Note that $\hat{\phi}$ is considered constant in the ground state %(inducing a long-range order);  breaking the global gauge symmetry. Performing the SW %map on Eq. (\ref{ourJ1}), we obtain
%$\label{ourJ1}
%\hat{\textbf{J}}_{s}=\dfrac{(2e)^{2}}{2m^{2}}{n}_{s}\hat{\textbf{A}}.$
%Comparando con la expresión Eq. (\ref{eq: NC London-1}) podemos notar que las %correciónes en $\theta$ a primer orden difieren por un factor cosntante, esto es por que %en nuestro trabajo seguimos el procedimiento de Drude postulando la ley de Ohm NC Eq. %(\ref{eq:noncommutative Ohm´s law}) y en este ultimo aproach obtuivimos la ec London a %partir de la variación de la energía libre Eq. (\ref{LEnergy}), sin embargo en el límite %$\theta=0$ en ambos casos recuperamos el caso usual.}
from the above equation, we will not expect to obtain an equivalent $\theta$-expansions relative to the ones shown in this work. The reason for this discrepancy is due to our treatment using as starting point, a constitutive equation, i.e., Ohm's law. 

To describe a general NCGL, the SW map which permits extracting the effects of NC, can be viewed as a mechanism to explore the nature of noncommutative geometry but likewise one can write it from different perspectives\cite{Gouba:2016iar}. In that sense, this work represents a first approach toward the development of a general NC Ginzburg-Landau field theory.

Moreover, a general NCGL theory with dynamical order parameter can be taken as
\begin{align}
    &\hat{F}_{s}(T) =\hat{F}_{n}(T)\nonumber \\
    &+\int\left(\dfrac{\hbar^{2}}{2m^{*}}|\hat{D}_{\mu}*\hat{\Psi}|^{2}+a_{eff}\Bar{\hat{\Psi}}*\hat{\Psi}+\dfrac{b_{eff}}{2}\Bar{\hat{\Psi}}*\hat{\Psi}*\Bar{\hat{\Psi}}*\hat{\Psi}- \frac{1}{4\mu_0}\hat{F}_{\mu\nu}*\hat{F}^{\mu\nu} \right)dr^{3}.
\end{align}
%and the free energy for the magnetic field given by $\hat{F}_{B}\sim - \int\left(\frac{1}{4\mu_0}\hat{F}_{\mu\nu}\hat{F}^{\mu\nu}\right)dr^{3}$.
Then, the NC supercurrent can be found with functional derivatives
\begin{equation}
    \hat{\textbf{J}}_{s}=-\dfrac{\delta \hat{F}_{s}}{\delta \hat{\textbf{A}}}=-\dfrac{2ie\hbar}{2m^{*}}\left(\Bar{\hat{\Psi}}*\nabla *\hat{\Psi}-\hat{\Psi}*\nabla*\Bar{\hat{\Psi}}\right)-\dfrac{4e^{2}}{m^{*}}\Bar{\hat{\Psi}}*\hat{\Psi}*\hat{\textbf{A}}.
\end{equation}
In order to perform our SW map, we need take to account that $\hat{n}_{s}\sim\hat{\bar{\Psi}}*\hat{\Psi}$ is a complex number that generally depends on $(\Lambda, \theta, \hat{A})$ due to coordinate dependency of the gauge parameter and NC vector potential. A $\theta$-expansion in the scalar field has the form \cite{Ulker:2007fm}
\begin{equation}
\widehat{\Psi}=\Psi-\frac{1}{4}\theta^{\mu\nu}\{A_{\mu},\,\left(D_{\nu}+\partial_{\nu}\right)\Psi\}+\mathcal{O}(\theta^{2}),
\end{equation}
hence, we need to settle down a suitable free energy functional that allows us to extract the NC current from a variational approach and perform SW mappings treating the gauge parameter $\Lambda$ in a consistent way. At the end of the day, such a gauge parameter is affected in the spontaneous symmetry breaking mechanism\footnote{Recall that $U_{*}(1)=e^{i\hat{\varLambda}(x)}$.}.

Another interesting construction in this direction is that the non-locality nature of the NC geometry translates into (or is equivalent to) spread out the sources, see for instance \cite{Banerjee:2009xx,Banerjee:2009gr} . With this in mind, an ingenious proposal define the free energy functional \cite{KOYAMA2013100} as
\begin{equation}
    F=\int d\textbf{r}\int d\textbf{r}'\left[ \mathcal{K}\cdot|\dfrac{1}{2m^{*}}\left(i\hbar\nabla-q\textbf{A}\right)\Psi|^{2}+\alpha|\Psi|^{2}+b|\Psi |^{4}+\dfrac{\textbf{B}^{2}}{8\pi} \right],
\end{equation}
being $\mathcal{K}$ some kernel that sum over nearby points $|\textbf{r}-\textbf{r}'|$ the effects of the non-local vector potential, namely
\begin{equation}
    \mathcal{K}(\textbf{r}-\textbf{r}')=f(\textbf{r}-\textbf{r}')\exp\left[ -\dfrac{ie}{h}\int_{x}^{y} d\textbf{S}\cdot \textbf{A}(s)\right].\label{ncgl}
\end{equation}
The function $f(\textbf{r}-\textbf{r}')$ makes the gradient non-local and supposed to decay in a length $|x|\sim \xi$. Comparing the Eq. (\ref{ncgl}) with Pippard's non-local current \cite{1950RSPSA.203..210P}
\begin{equation}
    \textbf{J}(\textbf{r})=-\dfrac{3}{4\pi\mu_{0}\xi\lambda^{2}}\int d^{3}\textbf{r}' \dfrac{(\textbf{r}-\textbf{r}')\left[(\textbf{r}-\textbf{r}')\cdot\textbf{A}(\textbf{r}')\right] }{\left(\textbf{r}-\textbf{r}'\right)^{4}}\exp\left[-\dfrac{|\textbf{r}-\textbf{r}'|}{\xi}\right],
\end{equation}
the nature of the kernel in NC scheme must be defined in terms of the $\theta$-parameter\footnote{The exponential dependency in terms of $\theta$ parameter is a well known procedure in NC field theory, see \cite{Maceda:2019woa} of one of the present authors.}, $\theta\sim a^{2}\xi_{0}^{2}$, as we stated in the introduction section. Hence, we can write
\begin{equation}
\mathcal{\vec{K}}_{NC}=\exp\left[-\dfrac{|\textbf{r}-\textbf{r}'|}{\sqrt{\theta}a^{-1}}\right]\dfrac{(\textbf{r}-\textbf{r}')}{|\textbf{r}-\textbf{r}'|^{4}}(\textbf{r}-\textbf{r}').
\end{equation}
To conclude, we emphasize that these constructions of full NCGL theory that allow us to address the SW mappings are interesting developments in their own right. The authors are working now on the systematic and coherent construction of the energy functional in order to extend the present work beyond the London limit.

\section*{Acknowledgments}
The authors would like to thank the financial support from science institutions in Mexico. D. M. Carbajal acknowledge financial support from COMECYT through the research grant No.CAT2021-0212 and for the partial support through  the CONACyT Network Project No. 376127, M. de la Cruz thanks Alfredo Herrera Aguilar and Postdoc for Mexico grant No. 30563 under CONACYT project A1-S-38041. S. Patiño thanks to CONACYT for PhD grant. We are also grateful to Dr. Jesús Martínez Martínez for the help.

%%%%%%%%%%%%%%%%%%%%%%%%%%%%%%%%%%%%%%%%%%%%%%%%%%%%%%%%%%%%%%%%%%%%%%%%%

\appendix\label{appendix}
\section{Noncommutative Maxwell's equations}
\label{secc:6}

The noncommutative Maxwell's equations in vacuum
\begin{eqnarray}
D_{\mu}* \widehat{F}^{\nu\mu} & = & \mu_{0}\widehat{J}_e^{\nu},
\label{eq:"Amp=0000E8re's law"}
\\[4pt]
D_{\mu}*\mathcal{\widehat{F}}^{\nu\mu} & = & 0,
\label{eq:"Gauss's law"}
\\[4pt]
D_{\mu}*\widehat{J}^{\mu} & = & 0,
\label{eq:continuity equation}
\end{eqnarray}
was proposed in~\cite{Maceda:2016ety, Langvik:2011gb} . Here 
\begin{eqnarray}
D_{\mu} & := & \partial_{\mu}-ie[\widehat{A}_{\mu},\,\cdot]_{*},
\nonumber \\[4pt]
\widehat{F}_{\mu\nu} & := & \partial_{\mu}\widehat{A}_{\nu}-\partial_{\nu}\widehat{A}_{\mu}-ie[\widehat{A}_{\mu},\,\widehat{A}_{\nu}]_{*},\end{eqnarray}
are the noncommutative covariant derivative and corresponding field strength tensor respectively. The dual field strength tensor is $\mathcal{\widehat{F}}^{\mu\nu} := \frac{1}{2}\epsilon^{\mu\nu\gamma\delta}\widehat{F}_{\gamma\delta}$ and Eq.~(\ref{eq:continuity equation}) is the NC continuity equation. Ampere's and Gauss law follows from the equations  Eqs.~(\ref{eq:"Amp=0000E8re's law"}) and~(\ref{eq:"Gauss's law"}), considering the magnetic and electric fields from the NC field strength  $\ensuremath{\widehat{B}_{i}=\frac{1}{2}\epsilon_{ijk}\widehat{F}^{jk}}$
and $\ensuremath{\widehat{E}_{i}=\widehat{F}^{oi}}$.

In signature $(-,+,+,+)$; we shall choose the components of the NC gauge potential as $\widehat{A}^{\mu}=\left(\hat{\phi},\widehat{A}_{i}\right)$.
%*******************
\subsection{Ampère's law:} 

Ampère's law can be obtained from Eq.~(\ref{eq:"Amp=0000E8re's law"}) by choosing $\nu=i$, for $i=1,2,3$. We have 
\begin{equation}
 D_{\mu}*\widehat{F}^{i\mu} = \mu_{0}\widehat{J}_{e}^{i}.\label{ampere2}
 \end{equation}

%For a static solution, the electric field $E^i := F^{0i}=0$ vanishes, and all time-dependence is suppressed. In consequence, only the spatial components of Ampère's law provide non-trivial information.
Let us focus then on $D_{\mu}*\widehat{F}^{i\mu}$, $i=1,2,3$. We have

\begin{align*}
D_{\mu}*\widehat{F}^{i\mu} & =D_{0}*\widehat{F}^{i0}+D_{j}*\widehat{F}^{ij}\\
 & =\partial_{0}\widehat{F}^{i0}-i[\widehat{A}_{0},\,\widehat{F}^{i0}]_{*}+\partial_{j}\widehat{F}^{ij}-i[\widehat{A}_{j},\,\widehat{F}^{ij}]_{*}\\
 & =-\partial_{0}\widehat{E}_{i}+i[\widehat{A}_{0},\,\widehat{E}_{i}]_{*}+\epsilon^{ijk}\left(\partial_{j}\widehat{B}_{k}-i[\widehat{A}_{j},\,\widehat{B}_{k}]_{*}\right),
\end{align*}
where we have defined $\epsilon^{ijk}\widehat{B}_{k}:=\widehat{F}^{ij}$ and $\ensuremath{\widehat{E}_{i}}:=\widehat{F}^{0i}$. 

In a perturbative treatment, the noncommutative gauge potential and the field strength tensor components can be written  as a power series up to second order on $\theta,$ the
noncommutative parameter
\begin{align}
D_{\mu}*\widehat{F}^{i\mu} & =\epsilon^{ijk}\left(\partial_{j}B_{k}^{0}+\partial_{j}B_{k}^{1}+\partial_{j}B_{k}^{2}-i[A_{j}^{0}+A_{j}^{1}+A_{j}^{2},\,B_{k}^{0}+B_{k}^{1}+B_{k}^{2}]_{*}\right)\nonumber \\
 & -\partial_{0}\left(E_{i}^{0}+E_{i}^{1}+E_{i}^{2}\right)+i[A_{0}^{0}+A_{0}^{1}+A_{0}^{2},\,E_{i}^{0}+E_{i}^{1}+E_{i}^{2}]_{*}\nonumber \\
 & =\epsilon^{ijk}\left(\partial_{j}B_{k}^{0}+\partial_{j}B_{k}^{1}+\partial_{j}B_{k}^{2}+\theta^{pq}\partial_{p}A_{j}^{0}(\partial_{q}B_{k}^{0}+\partial_{q}B_{k}^{1})+\theta^{pq}\partial_{p}A_{j}^{1}\partial_{q}B_{k}^{0}\right)\nonumber \\
 & -\partial_{0}\left(E_{i}^{0}+E_{i}^{1}+E_{i}^{2}\right)-\theta^{pq}\partial_{p}A_{0}^{0}(\partial_{q}E_{i}^{0}+\partial_{q}E_{i}^{1})-\theta^{pq}\partial_{p}A_{0}^{1}\partial_{q}E_{i}^{0}+\mathcal{O}(\theta^{3}),\label{eq: Ley de Faraday}
\end{align}
being $\epsilon^{jkl}B_{l}^{n}:=F_{n}^{jk}$ and $\ensuremath{E_{i}^{n}}:=F_{n}^{0i}$.

Using  the above result and writing electric current as a power series up to second order in $\theta$, i.e. $\widehat{J}_{e}^{i}=J_{e0}^{i}+J_{e1}^{i}+J_{e2}^{i}+\mathcal{O}(\theta^{3})$, the Eq.~(\ref{ampere2}) reads (S.I. units)
  % where $\widehat{J}_{e}^{i}=J_{e0}^{i}+J_{e1}^{i}+J_{e2}^{i}+\mathcal{O}(\theta^{3})$.  Notice that we are thus allowing the presence of an electric current into Maxwell's equations. It follows from Eq,~(\ref{ampere2}) that
\begin{align}
(\nabla\times\vec{B}^{0})^{i} & =\mu_{0}\varepsilon_{0}\partial_{0}E_{i}^{0}+\mu_{0}J_{e0}^{i},\nonumber \\
(\nabla\times\vec{B}^{1})^{i} & =\mu_{0}\varepsilon_{0}\partial_{0}E_{i}^{1}+\theta^{pq}\left(\epsilon^{ijk}\partial_{p}A_{j}^{0}\partial_{q}B_{k}^{0}-\mu_{0}\varepsilon_{0}\partial_{p}A_{0}^{0}\partial_{q}E_{i}^{0}\right)+\mu_{0}J_{e1}^{i},\nonumber \\
(\nabla\times\vec{B}^{2})^{i} & =\mu_{0}\varepsilon_{0}\partial_{0}E_{i}^{2}\nonumber\\
&+\theta^{pq}\left(\epsilon^{ijk}\partial_{p}A_{j}^{1}\partial_{q}B_{k}^{0}+\epsilon^{ijk}\partial_{p}A_{j}^{0}\partial_{q}B_{k}^{1}-\mu_{0}\varepsilon_{0}(\partial_{p}A_{0}^{1}\partial_{q}E_{i}^{0}+\partial_{p}A_{0}^{0}\partial_{q}E_{i}^{1})\right)\nonumber\\
&+\mu_{0}J_{e2}^{i},\label{eq:eq:Amp=00003D0000E8re's law zero, first and second order}
\end{align}
with $i,\,j\,,k=1,\,2,\,3$.
%*****************
\subsection{Electric Gauss's law:} 
Electric Gauss's law can be obtained from Eq.~(\ref{eq:"Amp=0000E8re's law"}) by choosing $\nu=0$ 
\begin{equation}
 D_{i}*\widehat{F}^{0i} = \mu_{0}\widehat{J}_{e}^{0},\label{Electric Gauss's law}
 \end{equation} 

where, $\widehat{J}_{e}^{0} \equiv c\widehat{\rho}(r)=\rho^{0}(r)+  \rho^{1}(r)+\rho^{2}(r)+\mathcal{O}(\theta^{3})$.
Following the same approach as the Ampere's law
%for the $i$-component
\begin{eqnarray}
&&\nabla\cdot\vec{E}^{0}=\frac{\rho^{0}(r)}{\varepsilon_{0}}, 
\nonumber \\[4pt]
&&\nabla\cdot\vec{E}^{1}=-\theta^{pq}\partial_{p}A_{i}^{0}\partial_{q}E_{i}^{0}+\frac{\rho^{1}(r)}{\varepsilon_{0}},
\nonumber \\[4pt]
&&\nabla\cdot\vec{E}^{2}=-\theta^{pq}\left(\partial_{p}A_{i}^{1}\partial_{q}E_{i}^{0}+\partial_{p}A_{i}^{0}\partial_{q}E_{i}^{1}\right)+\frac{\rho^{2}(r)}{\varepsilon_{0}}.
\label{eq:Gauss's law zero, first and second order}
\end{eqnarray}

%*
\subsection{Faraday's law:} 
Farday's law can follow from Eq.~(\ref{eq:"Gauss's law"}) by choosing $\nu=i$, for $i=1,2,3$, 
\begin{equation}
D_{\mu}*\mathcal{\widehat{F}}^{i\mu}=-\frac{1}{2}\epsilon^{ijk}D_{0}*\widehat{F}_{jk}+\epsilon^{ijk}D_{j}*\widehat{F}_{0k}=0.
\label{eq:"Faraday's law"}
 \end{equation} 
By second order $\theta$-expansion
% \begin{align}
%\frac{1}{2}\epsilon^{ijk}D_{0}*\widehat{F}_{jk} & %=\partial_{0}\left(B_{i}^{0}+B_{i}^{1}+B_{i}^{2}\right)+\theta^{pq}\partial_{p}A_{0}^{0}\left(\partial_{q}%B_{i}^{0}+\partial_{q}B_{i}^{1}\right)+\theta^{pq}\partial_{p}A_{0}^{1}\partial_{q}B_{i}^{0},\label{eq:Far%aday A}\\
%-D_{j}*\widehat{F}_{0k} & =\partial_{j}\left(E_{k}^{0}+E_{k}^{1}+E_{k}^{2}\right)+\theta^{pq}\partial_{p}A%_{j}^{0}\left(\partial_{q}E_{k}^{0}+\partial_{q}E_{k}^{1}\right)+\theta^{pq}\partial_{p}A_{j}^{1}\partial_%{q}E_{k}^{0}.\label{eq:Faraday B}
%\end{align}
%Using Eqs.~(\ref{eq:Faraday A}) and ~(\ref{eq:Faraday B}) into Eq.~(\ref{eq:"Faraday's law"}), we can write Faraday’s law order by order as

\begin{align}
(\nabla\times\vec{E}^{0})^{i} & =-\partial_{0}B_{i}^{0},\nonumber \\
(\nabla\times\vec{E}^{1})^{i} & =-\partial_{0}B_{i}^{1}-\theta^{pq}\left(\epsilon^{ijk}\partial_{p}A_{j}^{0}\partial_{q}E_{k}^{0}+\partial_{p}A_{0}^{0}\partial_{q}B_{i}^{0}\right),\nonumber \\
(\nabla\times\vec{E}^{2})^{i} & =-\partial_{0}B_{i}^{2}-\theta^{pq}\left(\epsilon^{ijk}\partial_{p}A_{j}^{1}\partial_{q}E_{k}^{0}+\epsilon^{ijk}\partial_{p}A_{j}^{0}\partial_{q}E_{k}^{1}+\partial_{p}A_{0}^{1}\partial_{q}B_{i}^{0}+\partial_{p}A_{0}^{0}\partial_{q}B_{i}^{1}\right).\label{eq: Faraday's law zero, first and second order}
\end{align}
%*****************
\subsection{ Magnetic Gauss's law:} 
for the sake of completeness, second order $\theta$-expansions for Gauss' law read
\begin{eqnarray}
&&\nabla\cdot\vec{B}^{0}=0, 
\nonumber \\[4pt]
&&\nabla\cdot\vec{B}^{1}=-\theta^{pq}\partial_{p}A_{i}^{0}\partial_{q}B_{i}^{0},
\nonumber \\[4pt]
&&\nabla\cdot\vec{B}^{2}=-\theta^{pq}\left(\partial_{p}A_{i}^{1}\partial_{q}B_{i}^{0}+\partial_{p}A_{i}^{0}\partial_{q}B_{i}^{1}\right).
\label{eq:Magnetic Gauss's law zero, first and second order}
\end{eqnarray}
First and second-order $\theta$-expansions allow the monopole behavior for a suitable choice of gauge potentials, as can be seen from the last two equations.

%%%%%%%%%%%%%%%%%%% BIBLIOGRAPHY %%%%%%%%%%%%%%%%%%%
\bibliography{Protocolo}

\bibliographystyle{./utphys}
\end{document}